\documentclass[journal=jpca,manuscript=article]{achemso}

\usepackage{achemso}
\usepackage{graphicx}
\usepackage{rotating}
\usepackage{amsmath}
\usepackage{amssymb}
\usepackage{color}

\renewcommand{\vec}[1]{{\bf #1}}

\SectionNumbersOn

\setkeys{acs}{articletitle=true}

\graphicspath{
  {./}
}

\title{The moving-grid effect in the harmonic vibrational frequency calculations with numeric atom-centered orbitals}
\author{Honghui Shang}
  \email{shanghui.ustc@gmail.com}
  \affiliation{State Key Laboratory of Computer Architecture, Institute of Computing Technology, Chinese Academy of Sciences, Beijing, 100190, China}

\author{Jinlong Yang}
\affiliation{Hefei National Laboratory for Physical Sciences at Microscale, Department of Chemical Physics, and Synergetic Innovation Center of Quantum Information and Quantum Physics, University of Science and Technology of China, Hefei, Anhui 230026, China}

\begin{document}

%%%%%%%%%%%%%%%%%%%%%%%%%%%%%%%%%%%%%%%%%%%%%%%%%%%%%%%%%%%%%%%%%%
%                            Abstract                            %
%%%%%%%%%%%%%%%%%%%%%%%%%%%%%%%%%%%%%%%%%%%%%%%%%%%%%%%%%%%%%%%%%%
\begin{abstract}
When using atom-centered integration grids, the portion of the grid that belongs to a certain atom also moves when this atom is displaced. In the paper, we investigate the moving-grid effect in the calculation of the harmonic vibrational frequencies when using all-electron full-potential numeric atomic-centered orbitals as the basis set. We find that, unlike the first order derivative (i.e., forces), the moving-grid effect plays an essential role for the second order derivatives (i.e., vibrational frequencies). Further analysis reveals that predominantly diagonal force constant terms are affected, which can be bypassed efficiently by invoking translational symmetry. Our approaches  have been demonstrated in both finite (molecules) and extended (periodic) systems. 
\end{abstract}

%
%\begin{keyword}
%vibrational frequencies\sep phonon band structure \sep atom-centered integration grids \sep numeric atomic orbitals\sep all-electron 
%%\MSC[2010] 00-01\sep  99-00
%\end{keyword}

\maketitle

%%%%%%%%%%%%%%%%%%%%%%%%%%%%%%%%%%%%%%%%%%%%%%%%%%%%%%%%%%%%%%%%%%
%                           Introduction                         %
%%%%%%%%%%%%%%%%%%%%%%%%%%%%%%%%%%%%%%%%%%%%%%%%%%%%%%%%%%%%%%%%%%

\section{Introduction}
%1.DFT 
Density-functional theory (DFT) \cite{Hohenberg1964,Kohn1965} has been developed into a widely applied ground-state method for polyatomic systems in chemistry, physics and material science. Besides, the response properties (e.g., polarizability, vibrational frequencies or phonon
 dispersions) related to the derivatives of the total energy can be calculated within the same quantum mechanical framework by
means of density-functional perturbation theory (DFPT) 
\cite{Gonze1997-1,Gonze1997-2,Baroni-2001} or so-called coupled perturbed 
self-consistent field
(CPSCF)method\cite{Gerratt-1967,Pople-1979,Dykstra-1984,Frisch-1990,
Ochsenfeld-1997, Liang-2005} in the quantum chemistry community. 
The popularizing of DFT in the quantum chemistry community came from an excellent 
paper by Johnson, Gill and Pople\cite{Johnson1993a}, in which they systematically studied the performance (optimized geometries, dipole moments, vibrational frequencies and atomization 
energies) of several different density functionals. In the paper, they also mentioned that in the calculation of the exchange-correlation energy gradient, the positions of the grid points are a central feature in the definition of the numerical exchange-correlation energy, which can be integrated numerically
using different kinds of grids. One naturally choice is the uniform grid 
which has been used in a number of DFT packages, e.g., OCTOPUS\cite{Octopus2015}, SIESTA\cite{Soler2002}. 
Another kind of grid is the atom-centered grid, which is defined such that an atom's grid ``moves with'' a displacement of its nucleus. Such an atom-centered grid was first proposed by Satoko\cite{Satoko1981} and then developed by Becke\cite{Becke1988}.
The advantage of the atom-centered grid are three-fold, firstly it could 
treat the full-electron system where the integrand is dominated by cusps 
at atomic nuclei; secondly, multicenter Poisson's equation can be 
reduced to a set of independent one-center Poisson's equations. 
Thirdly, such atomic-center-partition
scheme can bypass the so-called egg-box effect\cite{Soler2002} as shown in 
uniform grids. Thanks to the above advantages, such atomic-center grids have been widely adopted in the implementation of DFT in quantum chemistry software since 1990s\cite{Delley1991,Johnson1993,Johnson1994}.

\begin{figure}
 \includegraphics[width=\columnwidth]{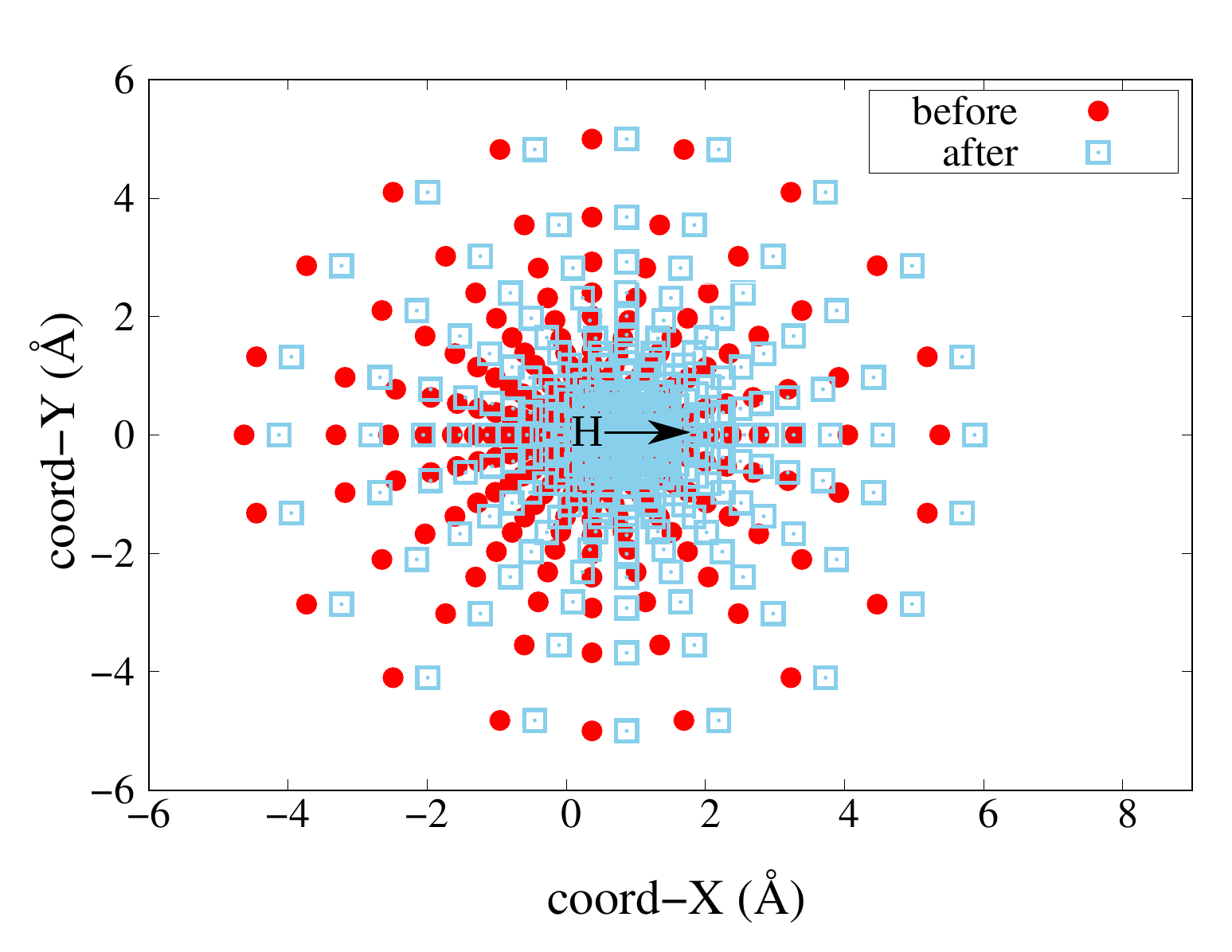}
 \caption{The coordinates (X and Y, with Z=0 \AA) of atom-centered grid for the hydrogen~(H) atom in the hydrogen molecule. The hydrogen atom is moved to the right side with 0.5 \AA as shown by arrow. Here
we show the atom-centered grids for H atom both before and after atom is moved.}
 \label{fig:moving_points}
\end{figure}

\begin{figure}
 \includegraphics[width=\columnwidth]{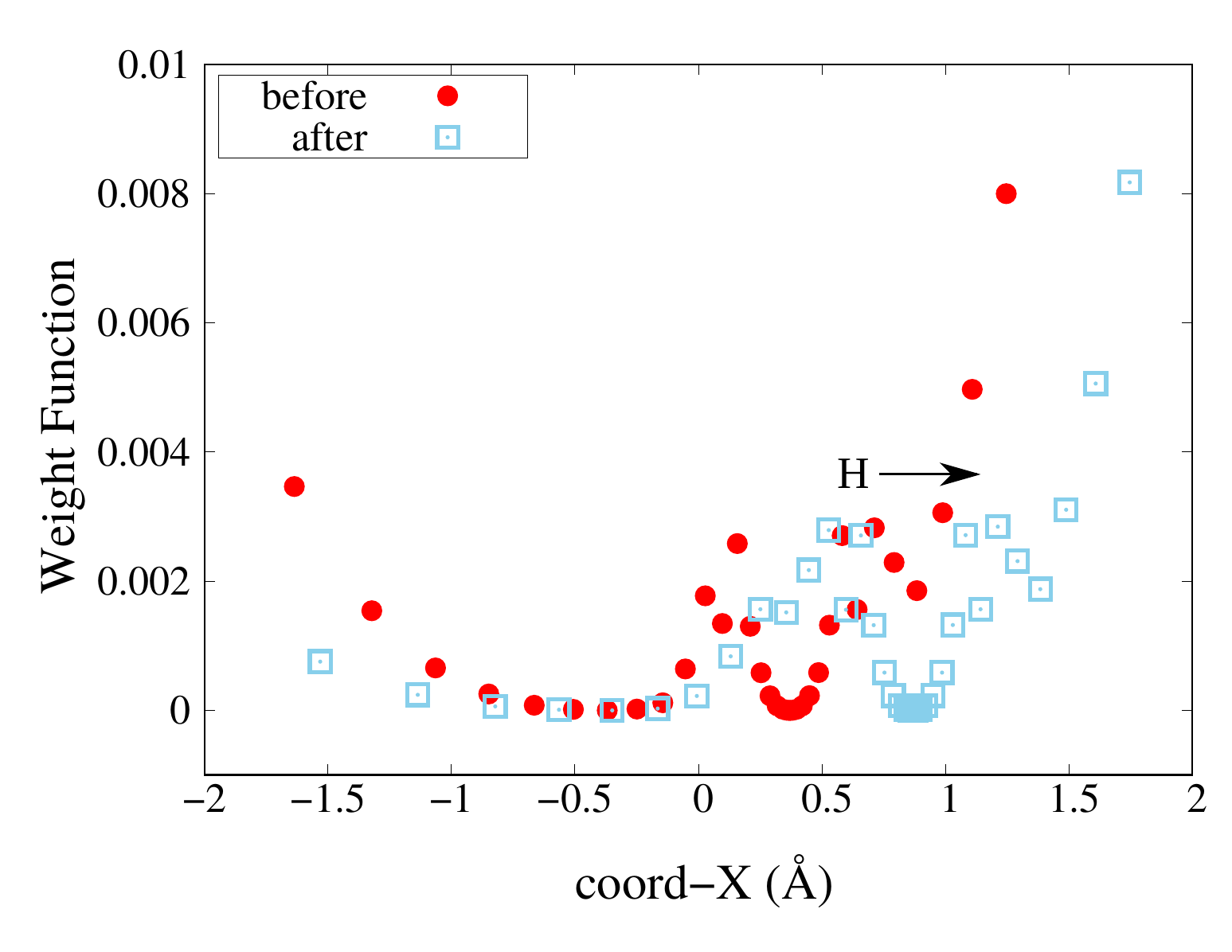}
 \caption{The weight function $w(\mathbf{r})$ (at Y=0 \AA and Z=0\AA) with respect to X coordinate for one hydrogen atom in the hydrogen molecule. The hydrogen atom is moved to the right side with 0.5 \AA ~ as shown by arrow. We could see the weight function of grids belong to the atom are changed.}
 \label{fig:weight_function}
\end{figure}

%3: moving-grid effect for Exc
However, this atomic partition scheme suffers from the so-called moving-grid effect
when derivatives are needed. This is because when an atom moves, all the points belong to this atom also move with it, for example, in Fig.~\ref{fig:moving_points}, the hydrogen atom labeled as H is moved to the right side, so the grids belong to this atom are also moved; In addition,
the integration weight functions are  also changed, so the derivatives of the weight function need to be included, as shown in Fig.~\ref{fig:weight_function}, 
when the hydrogen atom in the hydrogen molecule labeled as H is moved to the right side, the weight functions of this atom are also changed.
We call the above two phenomena (atom-centered grid points moving with the atom and the weight derivatives) as the moving-grid effect in this article. When the derivatives of the exchange-correlation energy are calculated using the Gaussian type orbitals~(GTO), this moving-grid effect is important for both force and Hessian only when the grids are of insufficient quality. \cite{Johnson1993a,Baker1994,Malagoli2003}.  It should be noted that although such moving-grid effect was called ``effect of quadrature weight derivatives'' in the previous literature\cite{Johnson1993,Baker1994,Malagoli2003}, in fact, they have already considered both the grid points and the weight derivatives in their work.

The moving-grid effect only appears for the numerical integrations. In the previous studies\cite{Johnson1993,Baker1994,Malagoli2003} with the Gaussian type orbitals~(GTOs), the moving-grid effect has been considered for force and Hessian calculations only for the exchange-correlation term. This is because when using the GTOs, only the exchange-correlation term is integrated numerically while the Hellmann-Feynman terms and the other Pulay terms are treated analytically with GTOs, so no moving-grid effect needs to be considered for these terms; However, when the numeric atom-centered orbitals~(NAOs) are adopted, both the Hellmann-Feynman and the whole Pulay terms are integrated numerically\cite{Delley1991}, so in principle,
the moving-grid effect using numeric basis set could be more serious than using the Gaussian basis set which treats Coulomb terms analytically. 
In 1991, Delley\cite{Delley1991} has made a first analytical force implementation using numeric basis set, in which he also mentioned that for first-order derivative calculation, the moving-grid effect only``results in a small residual  (e.g.,10$^{-3}$ a.u.) at the energy minimum'', which could be left out
in his opinion. However, how about its influence on the second-order derivatives~(force constants/Hessians)? To the best of our knowledge, until now, it is still unknown when using NAOs. As a result, the motivation of our current work is to see the influence of the moving-grid effect on both the Hellmann-Feynman Hessian and the Pulay Hessian terms when using the atom-centered grid together with the NAOs, which has not been examined before. 
Recently, we have completed our implementation\cite{Shang2017,Shang2018} of density-functional perturbation theory (DFPT) in the all-electron Fritz Haber Institut \textit{ab initio} molecular simulations (FHI-aims) package\citep{Blum2009}, especially for harmonic vibrational properties\cite{Shang2017} in molecules and solids, 
using the numeric atom-centered orbitals as basis functions. Here in this work, we find that the Hellmann-Feynman Hessian terms can not be improved by only increasing the integration grid quality as before, we have also carefully analyzed where the large errors come from, which has not been reported in the literature so far.

%5. our work

%6. outline
The remainder of the paper is organized as follows. Firstly, we will give the fundamental theoretical framework. Then, we validate our method and implementation for molecules and solid by comparing vibrational/phonon frequencies computed with DFPT to the ones computed via finite-differences or experimental results. Furthermore, we exhaustively investigate the convergence behavior with respect to the numerical parameters of the implementation~(basis set, integration grids, elements, etc.). Finally, we will summarize the main ideas and findings of this work.

\section{Method}
\label{sec:theory}

\subsection{Integration scheme}
\label{subsec:integration}

Most current numeric integration schemes for all-electron full-potential numeric atomic-centered orbitals stem from procedures described by Becke\cite{Becke1988} and Delley\cite{Delley-partition,Delley-aug}. Firstly, the grids are partitioned to single atoms using partition function defined\cite{Delley-partition,Blum2009} as 
\begin{equation}
p_{I}(\mathbf{r})=\dfrac{g_{I}(\mathbf{r})}{\sum_{I'}g_{I'}(\mathbf{r})} \,,
\end{equation}
here $g_{I}$ is the peaked function\cite{Delley-partition},
\begin{equation}
g_{I}(\mathbf{r})=\dfrac{n_{I}^{free}(|\mathbf{r}-\mathbf{R}_I|)}{|\mathbf{r}-\mathbf{R}_I|^2 } \,,
\end{equation}
where $n_{I}^{free}$ is the electron density of free atom. Our partition scheme is different from the Hirshfeld partitioning\cite{Hirshfeld1977}. In Hirshfeld partitioning, the partition function is calculated only with the density of the free-atoms~($n_{I}^{free}$), but here in our scheme, the partition function is achieved by using the peaked function~($g_{I}$), which is the free atom density divided by the distance square. Secondly, the single-center(atom) integration is further separated into radial and angular parts, that 
radially the atom-centered grid consists of several spherical integration shells with radial integration weight w$_{rad}$\cite{Blum2009,Baker1994}, and on these
shells, angular integration points are distributed in such a way that spherical harmonics up to a certain order are integrated exactly by using the Lebedev grids\cite{Delley-aug}, with angular integration weights
w$_{ang}$. 
Then the weight function $w_I(\mathbf{r})$ for grids belong to an atom can be written as:
\begin{equation}
w_I(\mathbf{r})=p_{I}(\mathbf{r})w_{rad}w_{ang} \,.
\end{equation}
With the above weight function, the integral K can be approximated by a discrete summation, 
\begin{eqnarray}
K   = & \int d\mathbf{r} f(\mathbf{r}, \mathbf{R} ) \nonumber \\
  \approx  & \sum_{I} \sum_{\mathbf{r}}  w_I(\mathbf{r})
f(\mathbf{r}, \mathbf{R}_{I}) \,,
\end{eqnarray}
Then the first-order derivative of the integrals is 
\begin{eqnarray}
\dfrac{d K }{d \mathbf{R}_{J} }  & 
 \approx & \sum_{I} \sum_{\mathbf{r}}
\underbrace{ \dfrac{\partial  w_I(\mathbf{r})}{\partial \mathbf{R}_{J}} }_{weight \ derivatives} f(\mathbf{r}, \mathbf{R}_{I})  \nonumber \\
&  + & \sum_{I} \sum_{\mathbf{r}} w_{I}(\mathbf{r})
\underbrace{ \dfrac{d  f(\mathbf{r}, \mathbf{R}_{I})}{d \mathbf{R}_{J}} }_{grid\ moving } \,.
\label{eq:int_derivative}
\end{eqnarray}
From the Eq.\ref{eq:int_derivative}, we can see that, both the weight derivatives and the atom-centered grid points moving with the atom are
need to be considered for the derivative calculations of the integrals.

\subsection{Force and force constants}
\label{sec:FC_fixed}
In DFT, the total energy is uniquely determined by the electron density $n(\mathbf{r})$  
\begin{eqnarray} 
  E_{tot}     & = & \underbrace{-\dfrac{1}{2}\sum_{i}<\phi_i|\nabla^2|\phi_i> }_{T_{s}[n]}     -  \underbrace{ \int {n(\mathbf{r})  \sum_{I}\dfrac{Z_{I}}{|\mathbf{r}-\mathbf{R}_{I}|}  d\mathbf{r}} }_{ E_{ext}[n]}  \nonumber   \\  
  & + & \underbrace{ \dfrac{1}{2}\int \int {\dfrac{n(\mathbf{r}) n(\mathbf{r'}) }{|\mathbf{r}-\mathbf{r'} |}  d\mathbf{r}  d\mathbf{r'}} }_{E_{H}[n]}  +   \underbrace{ \dfrac{1}{2}\sum_{I}\sum_{J}{\dfrac{Z_{I} Z_{J}}{|\mathbf{R}_{I}-\mathbf{R}_{J} |} }    }_{E_{ion-ion}} \nonumber   \\
  & + & E_{ext}[n] \;,
\label{eq:KSTOT}
\end{eqnarray}
in which $T_{s}$ is the kinetic energy of non-interacting electrons, $E_{ext}$ the electron-nuclear, $E_H$ the Hartree, $E_{xc}$ the exchange-correlation, and $E_{ion-ion}$ the ion-ion repulsion energy. All energies are functionals of the electron density. Here we avoid an explicitly spin-polarized notation, a formal generalization to collinear (scalar) spin-DFT is straightforward. 
The electron density is written with the eigenfunction,
\begin{equation}
n(\mathbf{r})=\sum_i f_i |\psi_i(r)|^2|,
\end{equation}
in which $f_i$ denotes the occupation number of eigenstate~$\psi_i$. Such
Kohn-Sham states can be expanded in a finite basis set~$\chi_\mu(\vec{r})$
\begin{equation}
\psi_i(\mathbf{r})=\sum_{\mu}C_{\mu i} \: \chi_{\mu}(\mathbf{r})\;,
\label{eq:expansion}
\end{equation}
using the expansion coefficients $C_{\mu i}$.

The force which is the first-order derivative of the total energy~($E_{tot}$) withe respect to the atomic displacement~($\mathbf{R}_{I}$) can be split into three terms 
\begin{equation}
\vec{F}_I = -\dfrac{d E_{tot}}{d \mathbf{R}_{I}} = \vec{F}_I^{HF} + \vec{F}_I^{P} + \vec{F}_I^{MP}\;.
\label{eq:forces}
\end{equation}
The Hellmann-Feynman term can be written as  
\begin{equation}
\vec{F}_I^{HF}=-\int {n(\mathbf{r})  \dfrac{Z_{I}(\mathbf{R}_{I}-\mathbf{r}  )  }{|\mathbf{r}-\mathbf{R}_{I}|^3}  d\mathbf{r}}  +
\sum_{J\ne I}{\dfrac{Z_{I} Z_{J} (\mathbf{R}_{I}-\mathbf{R}_{J})}{|\mathbf{R}_{I}-\mathbf{R}_{J} |^3} }\;.
\label{eq:F_HF_use} 
\end{equation}
where $n(\vec{r})$ is the electron density, $Z_{I}$ refer to the nuclear charge. 

The Pulay term can be written with Kohn-Sham Hamiltonian~($\hat{h}_{ks}= \hat{t}_s + \hat{v}_{ext}(r)+\hat{v}_{H}+\hat{v}_{xc}$), atomic basis set~$\chi_\mu(\vec{r})$, density matrix~($P_{\mu\nu}=\sum_{i}f_{i} C_{\mu i} C_{\nu i}$) and energy weighted density matrix~($W_{\mu\nu}=\sum_{i}f_{i}\varepsilon_i C_{\mu i} C_{\nu i}$ ), 
\begin{eqnarray}
\vec{F}_I^{P} = & 
- 2\sum_{\mu\nu} P_{\mu\nu}\int{  \dfrac{\partial \chi_{\mu}(\mathbf{r}) }{\partial \mathbf{R}_I} \hat{h}_{ks} \chi_{\nu}(\mathbf{r}) d \mathbf{r}  } \nonumber \\
 & +2\sum_{\mu\nu} W_{\mu\nu}\int { \dfrac{\partial \chi_{\mu}(\mathbf{r}) }{\partial \mathbf{R}_I} \chi_{\nu}(\mathbf{r}) d \mathbf{r} } \,,\label{eq:F_pulay_use} 
\end{eqnarray}
in which $\hat{t}_s$ is the single particle kinetic operator, $\hat{v}_{ext}$ the (external) electron-nuclear potential, $\hat{v}_{H}$ the Hartree potential, and $\hat{v}_{xc}$ the exchange-correlation potential, $\varepsilon_i$ is the eigenvalue, $f_i$ denotes the occupation number of eigenstate, $C_{\mu i}$ is the expansion coefficient. The multipole force~($\vec{F}_I^{MP}$) arising from the multipole correction\cite{Blum2009}.

Using the above force form, we get the force constants, the multipole term is omitted here  since its contribution is already three orders of magnitude smaller at the level of the forces. 
\begin{equation}
\Phi_{I,J}=\dfrac{d^2{E_{tot}} }{d{\mathbf{R}_{I}} d{\mathbf{R}_{J}}}=\Phi_{IJ}^{HF}+\Phi_{IJ}^{P} \;.
\label{eq:Hessian_KS}
\end{equation}
For the sake of readability, the total derivative of the Hellmann-Feynman term $\Phi_{I\alpha,J\beta}^{HF}$ is divided into two terms 
\begin{equation}
\Phi_{I,J}^{HF}=\Phi_{I,J}^{HF-r}+\Phi_{I, J}^{HF-R} \;.
\end{equation}
Then the first term which is the derivative of the 
first term of Eq.(\ref{eq:F_HF_use}), accounts for the response of the integration.
\begin{eqnarray}
\Phi_{I,J}^{HF-r} & =\int {  \dfrac{\partial{n(\mathbf{r})} }  
{ \partial{\mathbf{R}_{J}} } V_{I}^{(1)}(\mathbf{r})   d \mathbf{r} }  \nonumber \\ 
       & + \int { n(\mathbf{r}) \dfrac{\partial{ V_{I}^{(1)} } }  
{ \partial{\mathbf{R}_{J}} }   d \mathbf{r} }    \;.
\label{eq:Hessian_HF_r} 
\end{eqnarray}
If we label the first-order derivative of electron-ionic interaction as $V_{I}^{(1)}(\mathbf{r})$
\begin{equation}
V_{I}^{(1)}(\mathbf{r})=\dfrac{\partial \hat{v}_{ext} }{\partial \mathbf{R}_{I} } =  \dfrac{Z_{I}(\mathbf{R}_{I}-\mathbf{r})  }{|\mathbf{r}-\mathbf{R}_{I}|^3} \,,
\end{equation}
then the second order derivative of electron-ionic interaction is (with $\alpha$ and $\beta$ label the coordinates)
\begin{eqnarray}
 \dfrac{\partial{ V_{I\alpha}^{(1)}(\mathbf{r})  } }  
{ \partial{\mathbf{R}_{J\beta}} } & = \delta_{IJ}\delta_{\alpha \beta}  \dfrac{Z_I }{|\mathbf{r}-\mathbf{R}_{I}|^3 }  \nonumber   \\
   & -3\cdot \delta_{IJ}  \dfrac{Z_I(\mathbf{R}_{I}-\mathbf{r})_{\alpha}(\mathbf{R}_{J}-\mathbf{r})_{\beta} }{|\mathbf{r}-\mathbf{R}_{I}|^5 }  )  \;. 
 \label{eq:VI_deriv_fix}
\end{eqnarray}

The second term 
\begin{eqnarray}
\Phi_{I\alpha,J\beta}^{HF-R}= & (1-\delta_{IJ})\left[ \delta_{\alpha \beta} \dfrac{Z_I Z_J}{|\mathbf{R}_I-\mathbf{R}_{J}|^3 }   \right . \nonumber  \\ 
   & \left. - 3\cdot  \dfrac{Z_I Z_J (\mathbf{R}_{I}-\mathbf{R}_{J})_{\alpha}(\mathbf{R}_{I}-\mathbf{R}_{J})_{\beta} }{|\mathbf{R}_I-\mathbf{R}_{J}|^5 }   \right] \nonumber  \\ 
     & + \delta_{IJ}\left[-\delta_{\alpha \beta} \dfrac{Z_I Z_J}{|\mathbf{R}_I-\mathbf{R}_{J}|^3 }   \right . \nonumber  \\ 
   & \left. + 3 \cdot \dfrac{Z_I Z_J (\mathbf{R}_{I}-\mathbf{R}_{J})_{\alpha}(\mathbf{R}_{I}-\mathbf{R}_{J})_{\beta} }{|\mathbf{R}_I-\mathbf{R}_{J}|^5 }   \right]  \,, 
\label{eq:Hessian_HF_R} 
\end{eqnarray}
accounts for the response of the ionic-ionic summation.

Similarly, the total derivative of Pulay term $\Phi_{I,J}^{P}$ is split into four terms:
\begin{equation}
\Phi_{I,J}^{P} = \Phi_{I,J}^{P-P} + \Phi_{I,J}^{P-H} + \Phi_{I,J}^{P-W} + \Phi_{I,J}^{P-S}\;.
\end{equation}
The first term
\begin{equation}
\Phi_{I,J}^{P-P} =  {2} \sum_{\mu,\nu} 
\left( \frac{d P_{\mu,\nu}}{d \vec{R}_{J}}\right)
\int  \dfrac{\partial \chi_{\mu}(\mathbf{r})}{\partial  \mathbf{R}_{I} }\hat{h}_{KS} \chi_{\nu}(\mathbf{r}) \, d\mathbf{r} \,, 
\label{eq:PhiPP}
\end{equation}
accounts for the response of the density matrix~$P_{\mu,\nu }$. The second term 
\begin{eqnarray}
\Phi_{I,J}^{P-H} & = &  
2\sum_{\mu,\nu}  
P_{\mu,\nu} \cdot \nonumber \\ 
&&\left(
\int  \dfrac{\partial^2 \chi_{\mu }(\mathbf{r})}{\partial  \mathbf{R}_{I}\partial  \mathbf{R}_{J} }\,\hat{h}_{KS}\, \chi_{\nu}(\mathbf{r}) \, d\mathbf{r}\right. \nonumber \\
&&+\int  \dfrac{\partial \chi_{\mu}(\mathbf{r})}{\partial  \mathbf{R}_{I} }\frac{d \hat{h}_{KS}}{d\vec{R}_{J}}\chi_{\nu}(\mathbf{r}) \, d\mathbf{r} \nonumber \\
&&\left.+\int  \dfrac{\partial \chi_{\mu}(\mathbf{r})}{\partial  \mathbf{R}_{I} }\hat{h}_{KS}\frac{\partial \chi_{\nu}(\mathbf{r})}{\partial \vec{R}_{J}} \, d\mathbf{r}
\right) \,, 
\label{eq:PhiPH}
\end{eqnarray}
accounts for the response of the Hamiltonian~$\hat{h}_{ks}$, while the third  and fourth term
\begin{eqnarray}
\Phi_{I,J}^{P-W} & = &  - 2\sum_{\mu,\nu}
\frac{d W_{\mu,\nu}}{d\vec{R}_{J}}
\int  \dfrac{\partial \chi_{\mu}(\mathbf{r})}{\partial  \mathbf{R}_{I} }\chi_{\nu}(\mathbf{r}) \, d\mathbf{r}
\label{eq:PhiPW}\\
\Phi_{I,J}^{P-S} & = &  - 2\sum_{\mu,\nu } 
W_{\mu,\nu} \left(
\int  \dfrac{\partial^2 \chi_{\mu}(\mathbf{r})}{\partial  \mathbf{R}_{I} \partial  \mathbf{R}_{J}}\chi_{\nu}(\mathbf{r}) \right. \nonumber \\ 
                &   & + \left. \int  \dfrac{\partial \chi_{\mu}(\mathbf{r})}{\partial \mathbf{R}_{I} }\dfrac{\partial \chi_{\nu}(\mathbf{r}) }{\partial\vec{R}_{J}}  d\mathbf{r} \right) \,, 
\label{eq:PhiPS}
\end{eqnarray}
for the response of the energy weighted density matrix~$W_{\mu,\nu}$ and the overlap matrix~$S_{\mu,\nu}$, 
respectively.
Please note that in all four contributions many terms vanish due to the fact that 
we use localized atomic orbitals, 
\begin{equation}
\frac{\partial \chi_{\mu}(\vec{r})}{\partial \vec{R}_{J}} = 
\frac{\partial \chi_{\mu}(\vec{r})}{\partial \vec{R}_{J}} \delta_{I(\mu),J} \; .
\label{eq:basis_fixed}
\end{equation}
Similarly, it is important to
realize that all partial derivatives that appear in the force constants can be readily computed numerically, since the $\chi_{\mu m}$ are numeric atomic orbitals, which are defined using a spline radial function and spherical harmonics for the angular dependence~\cite{Blum2009}.

\subsection{Moving-grid effect in force constants calculation}
\label{sec:FC_moving}
In the force constants calculation, the moving-grid effect only appears in the terms that contain
integration. As noted by Baker\cite{Baker1994}, for force calculation,
essentially identical results could be obtained with the moderate size grids (around 3500 points per atom) whether or not the moving-grid effect is considered. In current work using FHI-aims, even the smallest grid sizes (light setting) is around 5000 grid points per atom(Tab.\ref{tab:H-grid}), so in current work, we have omitted the moving-grid effect in the force calculation~(first-order derivatives), and only focus on the moving-grid effect in the Hessian calculation~(second-order derivative). As a result, for the Hellmann-Feynman term, only $\Phi_{I\alpha,J\beta}^{HF-r}$ (Eq.\ref{eq:Hessian_HF_r}) need to be considered; For the Pulay term, only $\Phi_{I,J}^{P-H}$ (Eq.\ref{eq:PhiPH})and $\Phi_{I,J}^{P-S}$(Eq.\ref{eq:PhiPS}) are considered. We will show in detail in the following section. The Hessian for exchange-correlation part as discussed in the literature\cite{Johnson1993,Baker1994,Malagoli2003}  is already in the Pulay Hessian calculation. 

In the following, we will show our moving-grid scheme for the corresponding Hellmann-Feynman term and Pulay term. 

\subsubsection{Moving-grid effect in Hellmann-Feynman term } 
The $\Phi_{I,J}^{HF-r}$ (Eq.\ref{eq:Hessian_HF_r}) term is an integration, and we
need to use the approximation as shown in Eq.(\ref{eq:int_derivative}).

\begin{eqnarray}
\Phi_{I,J}^{HF-r} & \approx &  \sum_{\mathbf{r}} \underbrace{   \dfrac{\partial{ w(\mathbf{r} )} } { \partial{\mathbf{R}_{J}} } }_{weight \ derivatives}
n(\mathbf{r}) V_{I}^{(1)}{(\mathbf{r})}  \nonumber  \\ & + &  \sum_{\mathbf{r}}  w(\mathbf{r} )  \underbrace{   \left[ \dfrac{\partial{n(\mathbf{r})} }  
{ \partial{\mathbf{R}_{J}} }V_{I}^{(1)}{(\mathbf{r})} +
  n(\mathbf{r}) \dfrac{\partial{V_{I}^{(1)}
{(\mathbf{r}} )  }}{ \partial{\mathbf{R}_{J}} }     \right] }_{grid\ moving} \;. 
\label{eq:Hessian_HF_r_moving}
\end{eqnarray}
Here the first order density is
\begin{eqnarray}
\dfrac{\partial n(\mathbf{r})}{\partial \mathbf{R}_J}& = \sum_{\mu \nu} \dfrac{\partial  P_{\mu \nu}}{\partial \vec{R}_{J}} \chi_{\mu}(\vec{r})\chi_{\nu}(\vec{r}) \nonumber \\
   &+ \sum_{\mu \nu} P_{\mu \nu} \dfrac{\partial \chi_{\mu}(\vec{r})}{\partial \vec{R}_{J}} \chi_{\nu}(\vec{r}) \nonumber \\
    &+ \sum_{\mu \nu} P_{\mu \nu} \chi_{\mu}(\vec{r}) \dfrac{\partial \chi_{\nu}(\vec{r})}{\partial \vec{R}_{J}}  \;. 
\end{eqnarray}
When considering moving-grid effect, the derivative of basis function is written as:
\begin{equation}
\frac{\partial \chi_{\mu}(\vec{r})}{\partial \vec{R}_{J}} = \left\{ \begin{array}{rl}
 -\triangledown \chi_{\mu}(\vec{r})\delta_{I(\mu),J} &\mbox{ if $\mathbf{r} \notin atom_{J}$ } \\
  \triangledown \chi_{\mu}(\vec{r})(1-\delta_{I(\mu),J}) &\mbox{ if $\mathbf{r} \in atom_{J}$}
       \end{array} \right.
\end{equation}
and the electron-ionic interaction is 
\begin{equation}
\frac{\partial V_{I}^{(1)}(\mathbf{r})  }{\partial \vec{R}_{J}} = \left\{ \begin{array}{rl}
  Eq.(\ref{eq:VI_deriv_fix}) &\mbox{ if $\mathbf{r} \notin atom_{J}$ } \\
  Eq.(\ref{eq:VI_deriv_moving})  &\mbox{ if $\mathbf{r} \in atom_{J}$}
      \end{array} \right.
\end{equation}
When $r$ belong to $atom_{J}$, then it is moving grid, and the derivative of electron-ionic interaction is 
\begin{eqnarray}
 \dfrac{\partial{ V_{I\alpha}^{(1)}(\mathbf{r})   } }  
{ \partial{\mathbf{R}_{J\beta}} } & = -(1-\delta_{IJ})\delta_{\alpha \beta}  \dfrac{Z_I }{|\mathbf{r}-\mathbf{R}_{I}|^3 }  \nonumber   \\
   & +3\cdot (1-\delta_{IJ})  \dfrac{Z_I(\mathbf{R}_{I}-\mathbf{r})_{\alpha}(\mathbf{R}_{J}-\mathbf{r})_{\beta} }{|\mathbf{r}-\mathbf{R}_{I}|^5 }  ) \;.   
 \label{eq:VI_deriv_moving}
\end{eqnarray}

\subsubsection{Moving-grid effect in Pulay term } 
Similarly, using Eq.(\ref{eq:int_derivative}),the $\Phi_{I,J}^{P-H}$ (Eq.\ref{eq:PhiPH}) term can be
written as
\begin{eqnarray}
\Phi_{I,J}^{P-H} & \approx &  
2 \sum_{\mu,\nu} \sum_{\mathbf{r}} 
 \underbrace{P_{\mu,\nu}\cdot     \dfrac{\partial w(\mathbf{r})}{\partial  \mathbf{R}_{J} }  }_{weight \ derivatives }
\left[\dfrac{\partial \chi_{\mu}(\mathbf{r})}{\partial  \mathbf{R}_{I} }\hat{h}_{KS} \chi_{\nu}(\mathbf{r}) \right]
 \nonumber \\
&+&2\sum_{\mu,\nu} \sum_{\mathbf{r}} 
P_{\mu,\nu} \cdot w(\mathbf{r}) \nonumber \\ 
&\cdot&\left[
 \dfrac{\partial^2 \chi_{\mu }(\mathbf{r})}{\partial  \mathbf{R}_{I}\partial  \mathbf{R}_{J} }\,\hat{h}_{KS}\, \chi_{\nu}(\mathbf{r}) \, \right. \nonumber \\
&+&  \dfrac{\partial \chi_{\mu}(\mathbf{r})}{\partial  \mathbf{R}_{I} }\frac{d \hat{h}_{KS}}{d\vec{R}_{J}}\chi_{\nu}(\mathbf{r}) \,  \nonumber \\
&+&\underbrace{\left.  \dfrac{\partial \chi_{\mu}(\mathbf{r})}{\partial  \mathbf{R}_{I} }\hat{h}_{KS}\frac{\partial \chi_{\nu}(\mathbf{r})}{\partial \vec{R}_{J}} \,
\right] }_{grid\ moving}\;.  
\label{eq:PhiPH_moving}
\end{eqnarray}
In should be noted that, when using Gaussian basis set\cite{Johnson1993a,Baker1994,Malagoli2003},
only the Pulay term contains xc part need to be
considered, that the $\hat{h}_{KS}$ is replaced by $\hat{v}_{xc}$.

And finally, the integration form of $\Phi_{I,J}^{P-S}$(Eq.\ref{eq:PhiPS}) is
\begin{eqnarray}
\Phi_{I,J}^{P-S} & \approx &  
2 \sum_{\mu,\nu} \sum_{\mathbf{r}} 
\underbrace{W_{\mu,\nu}\cdot \dfrac{\partial w(\mathbf{r})}{\partial  \mathbf{R}_{J} } }_{weight \ derivatives}
\left[\dfrac{\partial \chi_{\mu}(\mathbf{r})}{\partial  \mathbf{R}_{I} } \chi_{\nu}(\mathbf{r}) \right]
 \nonumber \\
&+ &2\sum_{\mu,\nu} \sum_{\mathbf{r}} 
W_{\mu,\nu} \cdot w(\mathbf{r}) \nonumber \\ 
&\cdot&\underbrace{\left[
 \dfrac{\partial^2 \chi_{\mu }(\mathbf{r})}{\partial  \mathbf{R}_{I}\partial  \mathbf{R}_{J} }\chi_{\nu}(\mathbf{r}) + \dfrac{\partial \chi_{\mu}(\mathbf{r})}{\partial  \mathbf{R}_{I} }\frac{\partial \chi_{\nu}(\mathbf{r})}{\partial \vec{R}_{J}} \,
\right]}_{grid \ moving} \;.  
\label{eq:PhiPS_moving}
\end{eqnarray}
When considering moving-grid effect, the second derivative of basis function is written as,
\begin{equation}
\frac{\partial^2 \chi_{\mu}(\vec{r})}{\partial \vec{R}_{I}\vec{R}_{J}} = \left\{ \begin{array}{rl}
 \triangledown^2 \chi_{\mu}(\vec{r})\delta_{I(\mu),J} &\mbox{ if $\mathbf{r} \notin atom_{J}$ } \\
  -\triangledown^2 \chi_{\mu}(\vec{r})(1-\delta_{I(\mu),J}) &\mbox{ if $\mathbf{r} \in atom_{J}$}
       \end{array} \right.
       \label{eq:basis_moving}
\end{equation}

Other Pulay term $\Phi_{I,J}^{P-P}$(Eq.\ref{eq:PhiPP}) and  
$\Phi_{I,J}^{P-W}$(Eq.\ref{eq:PhiPW}) contains
the derivatives which similar to Pulay force, so
the moving-grid effect in these terms has not been considered as suggested by Delley\cite{Delley1991}.

\section{Results}
\label{sec:results}
We have investigated the validation of the moving-grid effect implementation for both the Hellmann-Feynman and the Pulay terms in Section \ref{subsec:validation} by comparing to the results obtained from finite-differences,  then we investigated the convergence of vibrational frequencies with respect to the numerical parameters in Section \ref{subsec:convergence}, and we extended the test to periodic system in Section \ref{subsec:phonon}.

\subsection{Validation against finite-differences}
\label{subsec:validation}
%---------moving-grid-effect vs grid-------------------
Before the validation,  we first introduce the atom-centered grid setting in FHI-aims~\cite{Blum2009}, which is charactered by light, tight and really-tight using different radial and angular integration points. For the 
radial part, the multiplier $N_{r,mult}$ determines the 
 spherical integration shells $N_r$, for example, $N_{r,mult}$=2 results in a total of $2N_{r} +1$ radial integration shells. For the  angular part, the angular integration points are distributed in such a way that spherical harmonics up to a certain order are integrated exactly by the use of Lebedev grids as proposed by Delley~\cite{Delley-aug}. Here, we characterize the angular integration grids by the maximum number of angular integration points~$N_{ang,max}$ used in the calculation. For hydrogen atom, the number of radial multiplier~($N_{r,mult}$), the maximum angular integration points~($N_{ang,max}$) as well as the total integration grids~($N$) are shown in Tab.\ref{tab:H-grid}.
Second, we present the basis set definition in FHI-aims~\cite{Blum2009}: a minimal basis includes the radial functions of the occupied orbitals of free atoms with noble gas configuration and quantum numbers of the additional valence functions,  and additional radial functions are added to make ``tier 1'' ,``tier 2'', and so on. This corresponds roughly to split-valence polarization basis using Gaussians, see Ref.\cite{Blum2009} for more details. Here in Table \ref{tab:H-grid_test_FC} and Table \ref{tab:H-grid_test_freq} we use tier 2 basis set. The local approximation for exchange and correlation~(LDA parametrization of Perdew and Zunger \cite{Perdew/Zunger:1981} for the correlation 
energy density of the homogeneous electron gas based on the data of Ceperley and 
Alder~\cite{Ceperley/Alder:1980}) is used in current work.

\begin{table}
\begin{tabular}{c|ccc}
\hline \hline
 Hydrogen atom  &   $N_{r,mult}$ & $N_{ang,max}$ & N  \\
             \hline          
light        &  1             & 302    & 4740    \\
tight        &  2             & 434    & 14450    \\   
really-tight &  2             & 590    & 19502   \\
\hline \hline
\end{tabular}
\caption{Different grid settings for the hydrogen atom. Here $N_{r,mult}$ refers to the number of radial multiplier, $N_{ang,max}$ refers to  the maximum angular integration points, $N$ refers to the total integration grids. }
\label{tab:H-grid}
\end{table} 

Then we begin the validation test. We first use the example of an helium atom to check the influence of the moving-grid effect on the Hellmann-Feynman Hessian~($\Phi^{HF}_{1x,1x}$) and the Pulay Hessian~($\Phi^{P}_{1x,1x}$). Obviously, the corresponding values should be zero since an atom does not exhibit any vibration mode. As shown in Table~\ref{tab:He-atom}, nonzero Hessian values always appeared if the moving-grid effect was ignored, regardless of whether the weight derivatives were considered or not
~(`no\_moving+no\_deriv'  or `no\_moving+yes\_deriv')
 It means that in the calculation of Hessian terms, the moving-grid effect should indeed be considered (`yes\_moving+no\_deriv'  or `yes\_moving+yes\_deriv'  ). It should be noted that the moving-grid effect in Pulay Hessian can be indeed reduced by increasing the integration grid quality, as shown in Table \ref{tab:He-atom}, the integration grid error for Pulay Hessian can be reduced from 1.6$\times$10$^{-4}$ to 2$\times$10$^{-5}$ by increasing the integration grid from light to tight.
This finding is just in agreement with Ref.\cite{Malagoli2003}, in which only the exchange-correlation part of the Pulay term has been considered. However, the Hellmann-Feynman Hessian term can not be improved by only increasing the integration grid quality,
as shown in Table \ref{tab:He-atom}, the Hellmann-Feynman Hessian is nearly a same nonzero number by increasing the integration grid from light to tight.
The large non-zero errors in the Hellmann-Feynman Hessian come from the inaccuracy calculation of the $1/\mathbf{r}^3$ potential term~(Eq.\ref{eq:VI_deriv_moving}) if neglecting the moving grid point, this is because the corresponding $1/\mathbf{r}^3$ integral 
is divergent. It should be stressed that this is not a problem for the Hellmann-Feynman force term~(Eq.\ref{eq:F_HF_use}) with $1/\mathbf{r}^2$ integral, whose
integrations errors can be reduced by increasing the integration grid quality.

\begin{table}
\begin{tabular}{c|ccc}
\hline \hline
light (N=5826) & $\Phi^{HF}_{1x,1x}$ & $\Phi^{P}_{1x,1x}$     & $\Phi^{Total}_{1x,1x}$     \\
\hline 
no\_moving+no\_deriv & -29.4 &  0.00016  &-29.4  \\
no\_moving+yes\_deriv & -29.4 &  0.00016  & -29.4 \\
yes\_moving+no\_deriv & 0.0  &  0.0  & 0.0  \\
yes\_moving+yes\_deriv & 0.0  &  0.0  & 0.0  \\
\hline
tight (N=16730)& $\Phi^{HF}_{1x,1x}$ & $\Phi^{P}_{1x,1x}$     & $\Phi^{Total}_{1x,1x}$     \\
\hline
no\_moving+no\_deriv & -29.5  &  -0.00002  &  -29.5 \\
no\_moving+yes\_deriv & -29.5  &  -0.00002  &  -29.5 \\
yes\_moving+no\_deriv & 0.0  &  0.0  & 0.0  \\
yes\_moving+yes\_deriv & 0.0   & 0.0 &  0.0  \\
\hline \hline
\end{tabular}
\caption{Hessian (Hartree/Bohr$^2$) of He atom computed with LDA functional, minimal basis set and different  numerical integration grids. }
\label{tab:He-atom}
\end{table} 

Then we present the Hessians/force constants results
for hydrogen dimer in Table \ref{tab:H-grid_test_FC}. In all cases, the force constants calculations were performed for the respective equilibrium geometry,~i.e.,~the structure obtained by relaxation~(maximum force~$<10^{-4}$~eV/$\mbox{\AA}$) using the exact the same computational settings. Here DFPT~(no\_moving+no\_deriv) means to neglect the moving grid point and the weight derivatives, by using Eq.(\ref{eq:Hessian_KS})-Eq.(\ref{eq:basis_fixed}) for all the force constants; DFPT~(no\_moving+yes\_deriv) means to only consider the weight derivatives and neglect the moving grid point; DFPT~(yes\_moving+no\_deriv) means to only consider the moving grid point and neglect the weight derivatives; DFPT~(yes\_moving+yes\_deriv) means to consider the  whole moving-grid effect, that both the moving grid point and the weight derivatives are considered by using Eq.(\ref{eq:Hessian_HF_r_moving})-Eq.(\ref{eq:basis_moving}). 
To validate with the above DFPT results, we have also obtained vibrational frequencies with finite-difference calculations, which was the first method to calculate the vibrational frequencies for molecules\cite{Handy1993,Malagoli2003}. The Hessian was obtained
via a first order finite difference expression for the forces using an atomic displacement of~0.0025~{\AA}. Here in finite-difference, we use Eq.(\ref{eq:F_HF_use}) for Hellmann-Feynman force. 

\begin{table*}
\scalebox{0.6}
{
\begin{tabular}{c  c  |   c | c |c | c| c}
\hline \hline  
  Hessian  &         &  DFPT     &  DFPT & DFPT & DFPT  &  finite-difference   \\
  (Hartree/Bohr$^2$)            &   & no\_moving+no\_deriv   & no\_moving+yes\_deriv  & yes\_moving+no\_deriv   & yes\_moving+yes\_deriv   \\  
\hline 
                     &  light     & -1.499  & -1.0411 & -0.1249    & 0.3331  & 0.3332\\ 
$\Phi^{HF}_{1x,1x}$  &  tight     & -1.499  &  -1.0427 & -0.01238    & 0.3332  & 0.3332   \\
                     &  really-tight & -1.499 & -1.0426 &  -0.1233   &  0.3333 & 0.3333 \\
\hline
                    &  light     & -0.3338  & -0.7919 &  0.1246   & -0.3334 &  -0.3332   \\ 
$\Phi^{HF}_{1x,2x}$  &  tight     & -0.3335 & -0.7906 &  0.1238  & -0.3332  & -0.3332 \\
             &  really-tight     & -0.3335 &  -0.7902 & 0.1233    & -0.3333  & -0.3333   \\
\hline             
                    &  light     & -0.00096  & 0.02968 & -0.03156   & -0.00092 & -0.001028    \\ 
$\Phi^{P}_{1x,1x}$  &  tight     &  -0.001184 & 0.02838 & -0.03076 & -0.001195  & -0.001185   \\
                &  really-tight  & -0.001188 & 0.02835 & -0.03073   & -0.001192  & -0.001189  \\
\hline 
                   &  light     &   0.001242 & -0.07426 & 0.07673  &0.001228   & 0.001028 \\ 
$\Phi^{P}_{1x,2x}$  &  tight     & 0.001197  &  -0.07449 & 0.07688  & 0.001197  & 0.001185   \\
                 &  really-tight &  0.001195 & -0.07449 & 0.07688    & 0.001194  & 0.001189  \\
%   \hline             
%                   &  light     &      &   &-0.0003   \\ 
%$\Phi^{MP}_{1x,1x}$  &  tight     &      &  &-0.0001    \\
%                   &  really-tight     &      &   &0.00001 \\
%\hline 
%                  &  light     &      &    & 0.0003   \\ 
%$\Phi^{MP}_{1x,2x}$  &  tight     &      &  & 0.0001   \\
%                  &  really-tight     &      & & -0.00001 \\    
                 \hline   \hline        
\end{tabular}
}
\caption{Hessian (Hartree/Bohr$^2$) of H$_2$ molecule computed with LDA functional, tier 2 basis set, and different  numerical integration grids. }
\label{tab:H-grid_test_FC}
\end{table*}

From Table \ref{tab:H-grid_test_FC}, we can see that 
(1) The diagonal part of the Hellmann-Feynman term in 
the force constant is wrong if omitting the moving-grid effect, no
matter which grids are using. For hydrogen, $\Phi^{HF}_{1x,1x}$ is not changed 
even really-tight setting grid is used. (2) The non-diagonal part of 
the Hellmann-Feynman term could be gotten using DFPT~(no\_moving+no\_deriv) scheme, by increasing the
grid, the relative error could be reduced from $0.18$\% (light) to  $0.06$\% (really-tight). (3) Compared with the Hellmann-Feynman term, the moving-grid in
Pulay term is smaller. As we use a large basis (tier 2) here, 
the Hellmann-Feynman term is $\sim 333$ times over Pulay term. And here for
hydrogen, the moving-grid effect seems very small for both diagonal part and
non-diagonal part in Pulay term. This is because the hydrogen is a light element,
and the moving-grid effect is not remarkable here, however in the following 
we will see the moving-grid effect is noticeable in the heavier elements.
(4) The DFPT~(no\_moving+yes\_deriv) results can not be
improved by increasing the grid quality for both the Hellmann-Feynman term and the Pulay term. Neither did the DFPT~(yes\_moving+no\_deriv) results. 
%(4) The percentage of multipole correction term in total force constant 
%is around $0.1 \%$ with light grid setting and less than $0.003 \%$ with really-tight grid setting, so it has been omitted in the following force constants calculation.   

\begin{table}
\begin{tabular}{c | c c  c  }
\hline \hline
  Frequency($cm^{-1}$)  & DFPT & finite-difference  & $\Delta$ \\  
  &  yes\_moving+yes\_deriv &   &  \\  
\hline 
    light  & 4173.6  & 4171.5 & 2.1  \\  
    tight &  4172.2  & 4171.5 & 0.7 \\  
  really-tight & 4172.9 & 4173.2 & 0.3  \\  
\hline \hline 
\end{tabular}
\caption{Frequencies (cm$^{-1}$) of H$_2$ molecule computed with LDA functional, tier 2 basis set.}
\label{tab:H-grid_test_freq}
\end{table}

Then for the vibrational frequencies test as shown in Table \ref{tab:H-grid_test_freq}, we can see that by considering
moving-grid effect~(yes\_moving+yes\_deriv), an excellent agreement with finite difference method is achieved. And  the absolute error reduced from $2.1~cm^{-1}$ with the light setting grid to $0.3~cm^{-1}$ with the really-tight grid. 

\begin{table}
\scalebox{0.8}
{
\begin{tabular}{ c c | c    c    c c    }
\hline \hline  
   &   Frequency($cm^{-1}$)    &  DFPT  &  DFPT & DFPT   & finite-difference    \\ 
   &                              &  fix-HF  &  fix-Pulay &  &    \\ 
\hline \hline  
   &    minimal       & -3683.6 &  3341.2  &   3341.5       &  3341.3 \\ 
H$_2$ & tier 1         & -5603.9 &  4206.6    &   4206.5      &  4207.1  \\    
   & tier 2         & -5533.9 & 4172.9    & 4172.9      &  4173.2 \\ 
\hline
   &  minimal         & -152460.1   &  976.5    &   969.5       & 967.9    \\ 
F$_2$ & tier 1         & -152431.1    & 1061.2     &  1054.8      & 1055.0    \\    
   & tier 2         & -411634.1   & 1072.1    &    1062.8      &  1063.2    \\ 
\hline
   &  minimal        & -411634.1   &  676.7   &  476.7      &   475.2    \\ 
Cl$_2$ & tier 1       & -411600.8    & 738.3   &    565.8     &   563.4   \\    
   & tier 2        & -411600.6   &  737.2   &    564.2    &     561.9  \\ 
\hline \hline 
\end{tabular}
}
\caption{The moving-grid effect with respect to different basis set and different element with really-tight grid and LDA functional. Here fix-HF refer to omit the moving-grid effect in the Hellmann-Feynman term;  fix-Pulay refer to omit the moving-grid effect in the Pulay term; DFPT means to consider the moving-grid effect~(yes\_moving+yes\_deriv) in both the Hellmann-Feynman and the Pulay terms. }
\label{tab:H2-F2-Cl2}
\end{table}

%---------moving-grid-effect vs element and basis-------------------
In order to see the influence of different basis set and elements, we present
harmonic frequencies for different dimers H$_2$, F$_2$ and Cl$_2$ computed with
LDA functional and really-tight grid setting using three different basis sets: minimal, tier1 and tier2 as described above. DFPT frequencies are computed analytically for three conditions:
(1) fix-HF: we omit the moving-grid effect in the Hellmann-Feynman terms, and only consider the moving-grid effect in the Pulay term; (2) fix-Pulay, we omit the moving-grid effect in the Pulay term, and only consider the moving-grid effect in the Hellmann-Feynman term; (3) DFPT, we consider
moving-grid effect~(yes\_moving+yes\_deriv) in both the Hellmann-Feynman and the Pulay terms. In finite-difference calculations, the force constants are also obtained via a first order finite difference expression for the forces using an atomic displacement of~0.0025~{\AA}, using Eq.(\ref{eq:F_HF_use}) for Hellmann-Feynman force.  It should be noted that, a larger atomic displacement of~0.013~{\AA} is also tested for the H$_2$, F$_2$ and Cl$_2$ systems, and we find that the relative errors are within 0.06\% compared with the results of 0.0025~{\AA} displacement.  A detailed list of results is given in the Table~\ref{tab:H2-F2-Cl2}. It can be seen that an excellent agreement between our DFPT implementation and the finite-difference results. The difference between frequencies computed using the DFPT-moving method and finite-difference method is typically less than 3 cm$^{-1}$, which is acceptable, (the largest absolute error 2.4 cm$^{-1}$ occurs for Cl$_2$ with tier 1 basis set); On the other hand, we could see the significant problem without considering moving-grid effect. (1) For DFPT-fix-Pulay, F$_2$ has errors of 7 cm$^{-1}$ (mini), 6.4 cm$^{-1}$ (tier 1) and 9.3 cm$^{-1}$(tier 2). The situation further worsens for 
Cl$_2$, which has errors of 200 cm$^{-1}$(mini),  172.5 cm$^{-1}$(tier 1) and 173.0 cm$^{-1}$(tier 2).  (2) For DFPT-fix-HF, all the frequencies just go to negative which is completely wrong. 

\begin{table}
\scalebox{0.5}
{
\begin{tabular}{c| c c c c }
\hline \hline
       &finite-difference &    DFPT&  ab-error  & rel-error\\
\hline
C$_2$H$_2$ & 637.90 & 642.33 & 4.43 & 0.69 \\
 & 723.46 & 727.19 & 3.73 & 0.52 \\
 & 723.46 & 727.19 & 3.73 & 0.52 \\
 & 2022.85 & 2021.80 & 1.05 & 0.05 \\
 & 3315.94 & 3312.56 & 3.38 & 0.10 \\
 & 3416.95 & 3413.87 & 3.08 & 0.09 \\
H$_2$CO & 1135.00 & 1134.66 & .34 & 0.03 \\
 & 1211.17 & 1211.75 & 0.58 & 0.05 \\
 & 1454.54 & 1455.90 & 1.36 & 0.09 \\
 & 1804.24 & 1802.20 & 2.04 & 0.11 \\
 & 2764.57 & 2764.09 & .48 & 0.02 \\
 & 2815.06 & 2814.57 & .49 & 0.02 \\
H$_2$O$_2$ & 350.82 & 350.92 & 0.10 & 0.03 \\
 & 959.96 & 960.16 & 0.20 & 0.02 \\
 & 1286.43 & 1285.05 & 1.38 & 0.11 \\
 & 1389.09 & 1389.18 & 0.09 & 0.01 \\
 & 3641.23 & 3640.56 & 0.67 & 0.02 \\
 & 3642.28 & 3642.07 & 0.21 & 0.01 \\
NH$_3$ & 931.60 & 933.16 & 1.56 & 0.17 \\
 & 1574.31 & 1575.66 & 1.35 & 0.09 \\
 & 1584.83 & 1584.09 & 0.74 & 0.05 \\
 & 3391.56 & 3391.27 & 0.29 & 0.01 \\
 & 3525.42 & 3524.85 & 0.57 & 0.02 \\
 & 3525.79 & 3525.10 & 0.69 & 0.02 \\
PH$_3$ & 944.29 & 944.40 & 0.11 & 0.01 \\
 & 1062.49 & 1066.17 & 3.68 & 0.35 \\
 & 1069.76 & 1070.79 & 1.03 & 0.10 \\
 & 2323.99 & 2323.49 & 0.50 & 0.02 \\
 & 2337.90 & 2338.45 & 0.55 & 0.02 \\
 & 2338.53 & 2338.69 & 0.16 & 0.01 \\
N$_2$H$_4$ & 468.98 & 468.80 & 0.18 & 0.04 \\
 & 701.29 & 702.52 & 1.23 & 0.18 \\
 & 865.90 & 866.78 & 0.88 & 0.10 \\
 & 1138.24 & 1137.91 & 0.33 & 0.03 \\
 & 1234.87 & 1235.03 & 0.16 & 0.01 \\
 & 1265.95 & 1264.61 & 1.34 & 0.11 \\
 & 1585.39 & 1587.87 & 2.48 & 0.16 \\
 & 1599.52 & 1598.81 & 0.71 & 0.04 \\
 & 3367.81 & 3367.44 & 0.37 & 0.01 \\
 & 3371.30 & 3370.66 & 0.64 & 0.02 \\
 & 3473.28 & 3472.58 & 0.70 & 0.02 \\
 & 3478.11 & 3477.45 & 0.66 & 0.02 \\ 
C$_2$H$_4$ & 791.76 & 792.27 & 0.51 & 0.06 \\
 & 915.72 & 914.13 & 1.59 & 0.17 \\
 & 935.45 & 933.53 & 1.92 & 0.21 \\
 & 1020.04 & 1020.92 & 0.88 & 0.09 \\
 & 1183.45 & 1184.04 & 0.59 & 0.05 \\
 & 1321.04 & 1320.68 & 0.36 & 0.03 \\
 & 1389.64 & 1391.67 & 2.03 & 0.15 \\
 & 1650.77 & 1650.20 & 0.57 & 0.03 \\
 & 3040.63 & 3040.78 & 0.15 & 0.00 \\
 & 3053.67 & 3053.24 & 0.43 & 0.01 \\
 & 3117.51 & 3117.60 & 0.09 & 0.00 \\
 & 3144.46 & 3143.90 & 0.56 & 0.02 \\
Si$_2$H$_6$ & 137.67 & 139.49 & 1.82 & 1.32 \\
& 319.19 & 321.68 & 2.49 & 0.78 \\
& 327.57 & 324.92 & 2.65 & 0.81 \\
& 429.19 & 429.72 & 0.53 & 0.12 \\
& 584.94 & 583.22 & 1.72 & 0.29 \\
& 590.59 & 587.19 & 3.40 & 0.58 \\
& 763.41 & 767.85 & 4.44 & 0.58 \\
& 839.46 & 836.22 & 3.24 & 0.39 \\
& 881.34 & 881.94 & 0.60 & 0.07 \\
& 883.10 & 883.18 & 0.08 & 0.01 \\
& 894.45 & 895.43 & 0.98 & 0.11 \\
& 898.27 & 897.42 & 0.85 & 0.09 \\
& 2145.32 & 2145.60 & 0.28 & 0.01 \\
& 2149.35 & 2149.51 & 0.16 & 0.01 \\
& 2159.21 & 2159.40 & 0.19 & 0.01 \\
& 2159.64 & 2159.50 & 0.14 & 0.01 \\
& 2168.67 & 2168.88 & 0.21 & 0.01 \\
& 2168.84 & 2168.99 & 0.15 & 0.01 \\ 
 \hline
MAE &        &          & 1.2 &    \\
MAPE&        &          &       & 0.13\%\\
\hline \hline
\end{tabular}
}
\caption{The comparison between the numerical and the analytical vibrational frequencies for 8 molecules. All calculations are performed at the LDA level of theory with fully converged numerical settings and relaxed geometries. DFPT means to consider the moving-grid effect~(yes\_moving+yes\_deriv).}
\label{tab:molecules test}
\end{table} 

To validate our DFPT implementation with moving-grid effect~(yes\_moving+yes\_deriv) in a more systematic way, we have also compared the vibrational frequencies of 8 selected molecules with finite-difference calculations.
In the finite-difference method, the force constants are calculated based on the first order derivative of the force, here using Eq.(\ref{eq:F_HF_use}) for Hellmann-Feynman force. 
Such numerical derivatives from finite-difference are quite dependent on the chosen displacement, which should not be too small nor too large in order to reduce the numerical error, and the value 0.013~{\AA} is just such intermediate displacement.
For example, in Si$_2$H$_6$, by increasing the atomic displacement from 0.0025~{\AA} to 0.013~{\AA} in the finite-difference method, the mean absolute error reduced from 2.4 cm$^{-1}$ to 1.3 cm$^{-1}$, however, by increasing the atomic displacement from 0.013~{\AA}  to 0.13~{\AA}, the mean absolute error increased from 1.3 cm$^{-1}$ to 19.1 cm$^{-1}$. 
So in the following, the atomic displacement of~0.013~{\AA} was used in the finite-difference calculations.
All calculations were performed at the LDA level of theory using fully converged numerical parameters : ``tier 2'' basis sets and ``really tight'' defaults were used for the numerical settings. Additionally, we increased the order of the multipole expansion to~$l=12$ and the radial integration grid to $N_{r,mult}=4$ for all systems.

In total, we find the  mean absolute error (MAE) is 1.2~cm$^{-1}$ and the mean absolute percentage error (MAPE) is $0.13$\%, which shows an excellent agreement between our DFPT implementation and the finite-difference results.
It should be noted that the largest occurring absolute error~(4.44~cm$^{-1}$ in S$_2$H$_6$) and the largest occurring relative error~($1.32$\% in Si$_2$H$_6$) still correspond to relatively moderate relative and absolute errors~($0.58$\% and $1.82$~cm$^{-1}$, respectively). 

As discussed for Table~\ref{tab:H-grid_test_FC}, the moving-grid effect has a much smaller influence on the non-diagonal terms of the force constant. This is because the form of the non-diagonal term is similar to the force calculation, which has been previously shown\cite{Baker1994} that the moving-grid effect could be neglected if a sufficient grid is used (large than 3500 per atom). As a result, we could use DFPT~(no\_moving+no\_deriv) method to get the non-diagonal terms as shown in Table~\ref{tab:H-grid_test_FC} and then using translational symmetry~(the system will keep invariance under any translation), which is also known as acoustic sum rule (ASR)\cite{Baroni-2001, Gonze1997-2,Deglmann2002} to get 
the diagonal terms:  
\begin{equation}
\dfrac{\partial^2 E_{tot}}{\partial R_{I\alpha} \partial R_{I\beta}}
=-\sum_{J\ne I}{\dfrac{\partial^2 E_{tot}}{\partial R_{I\alpha} \partial R_{J\beta}}  }\;,
\end{equation}
Such an ASR scheme has been implemented and adopted in Ref.\cite{Shang2017}, and an excellent agreement between this scheme with the finite-difference results for both molecules and solids systems has also been achieved\cite{Shang2017}. It should be noted that, when using the off-diagonal Hessian elements together with the ASR, the diagonal terms do not need to be calculated, so the computation time has been reduced ~(e.g 1.6\% time saving for  methane). We do not project out the rotational modes
from the force constants in this work since we also need to deal with the solid systems, which do not have the rotational invariance. In the following, we will use the DFPT+ASR method to make the convergence study.

\subsection{Convergence with respect to Numerical Parameters}
\label{subsec:convergence}
The convergence behavior of the DFPT method with respect to the numerical
parameters~(basis set size, numerical integration grids) is also analyzed. 
We use the four frequencies of methane~(CH$_4$) as the example, the DFPT calculations were performed for the respective equilibrium geometry,~i.e.,~the structure obtained by relaxation~(maximum force~$<10^{-4}$~eV/$\mbox{\AA}$) using the same computational settings. 

\begin{figure}
 \centering
 \includegraphics[width=0.9\columnwidth]{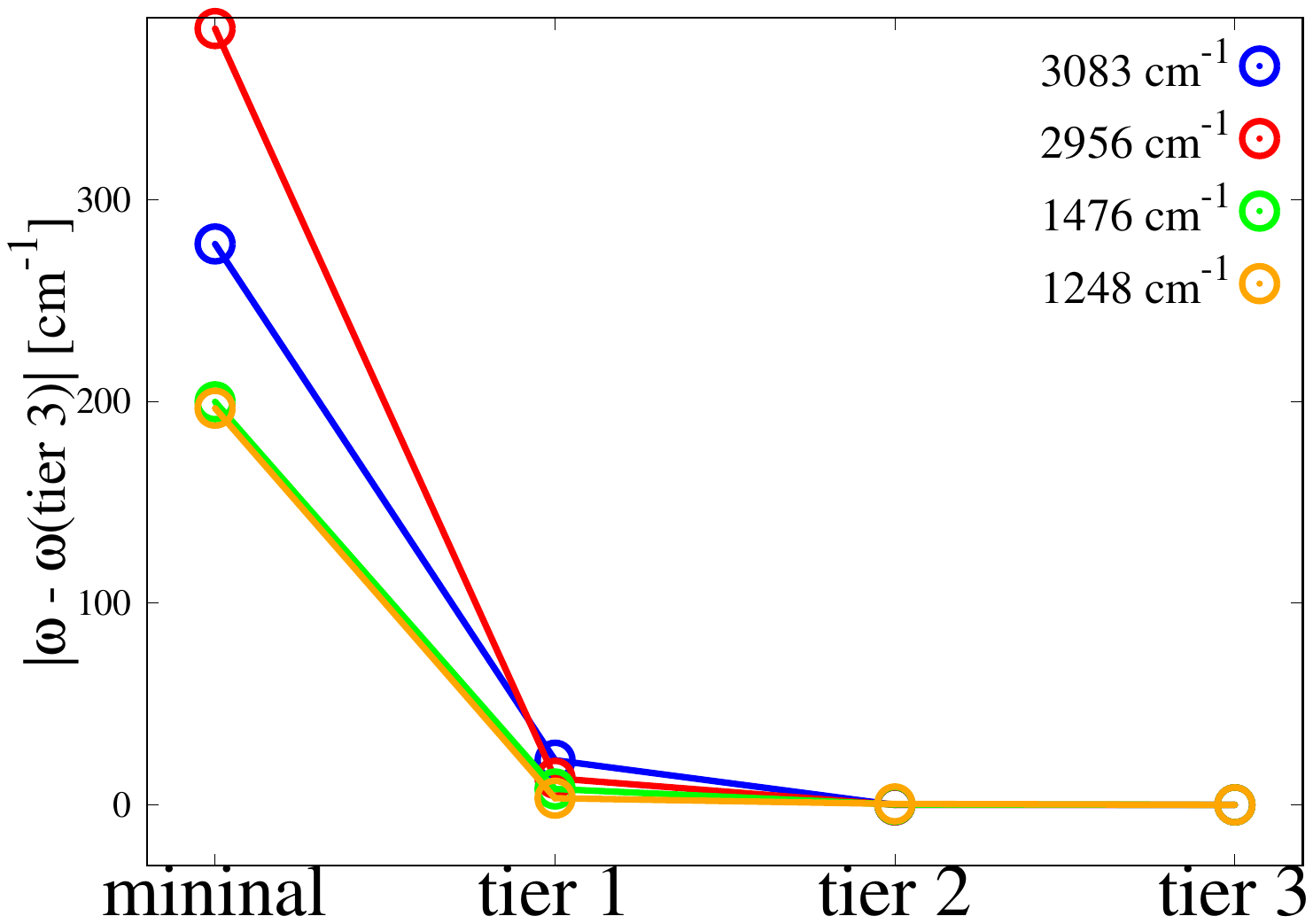}
 \caption{Convergence of the infrared-active vibrational frequencies of methane with respect to the basis set size~(see . We use really-tight grid setting with $N_{r,mult}$=4 and $N_{ang,max}$=590. The benchmark values are calculated using ``tier 3''.}
  \label{fig:ch4_basis}
\end{figure}

Fig.~\ref{fig:ch4_basis} shows the absolute change in these
vibrational frequencies if the basis set size is increased. The vibrational frequencies converge quickly with the basis set size. We get qualitatively correct results with a maximal absolute/relative error of $22$~cm$^{-1}$/$0.7$~\% at a ``tier 1'' level, and the results get fully quantitatively converged with the ``tier 2'' basis set with absolute and relative errors of $0.5$~cm$^{-1}$ and $0.04$~\%..

\begin{figure}
\centering
 \includegraphics[width=0.9\columnwidth]{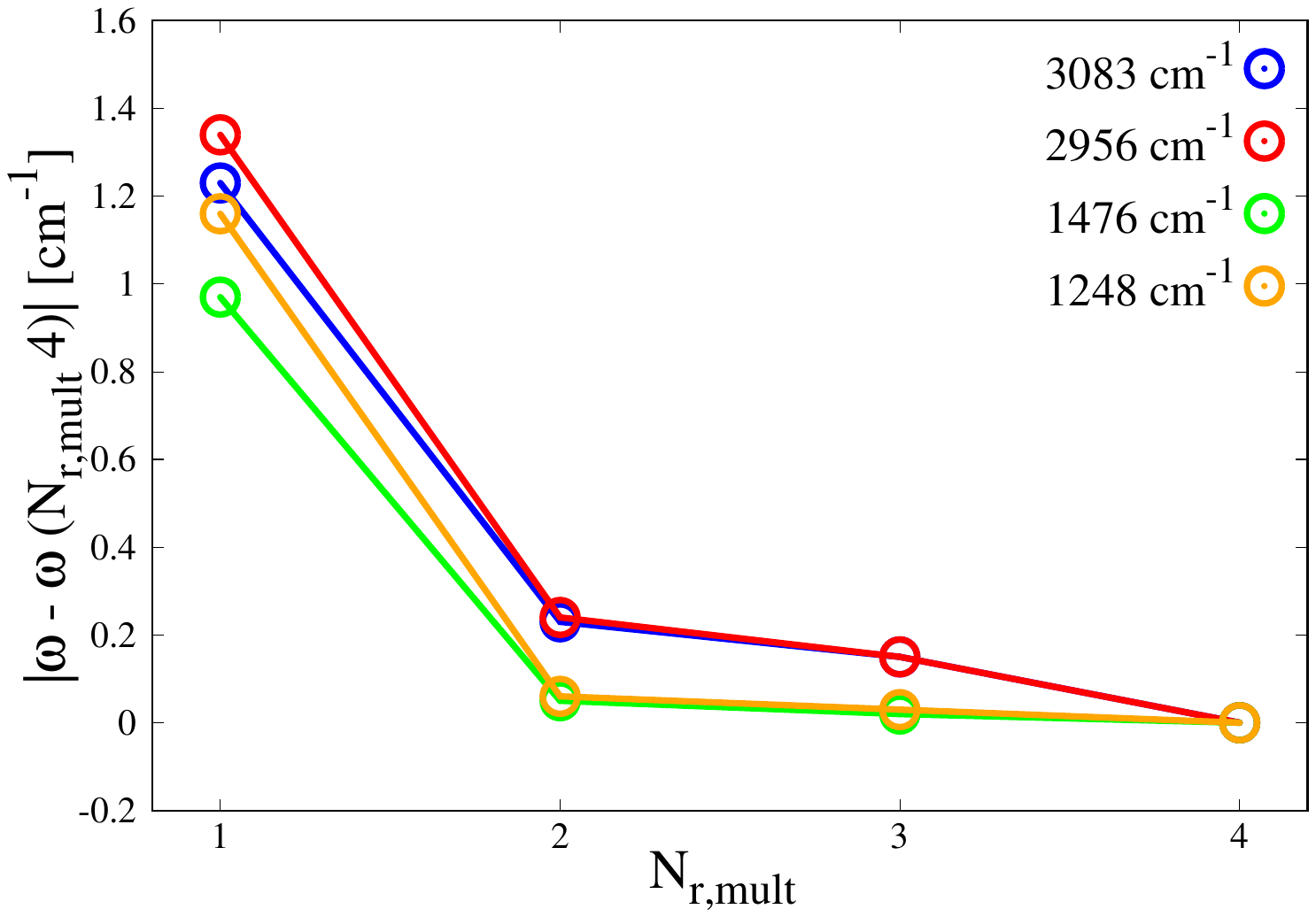}
 \caption{
 Convergence of the infrared-active vibrational frequencies of methane with respect to the radial grid density,
 as controlled by the parameter~$N_{r,mult}$~(see text). 
 We use a ``tier 2'' basis set and  $N_{ang,max}$=590 here. The benchmark values are calculated using $N_{r,mult}$=4.}
 \label{fig:ch4_Nr}
\end{figure}

Fig.~\ref{fig:ch4_Nr} shows our convergence tests with respect to the radial integration grids~($N_{r,mult}$), we find even the most sparse radial integrations grids yields qualitative and almost quantitatively correct frequencies, since at the $N_{r,mult}=1$ level the maximum absolute and relative errors are $1.3$~cm$^{-1}$ and $0.05$~\%. Quantitatively converged results are achieved at the $N_{r,mult}=2$ level with absolute and relative errors of $0.24$~cm$^{-1}$ and $0.01$~\%.

\begin{figure}
 \includegraphics[width=0.9\columnwidth]{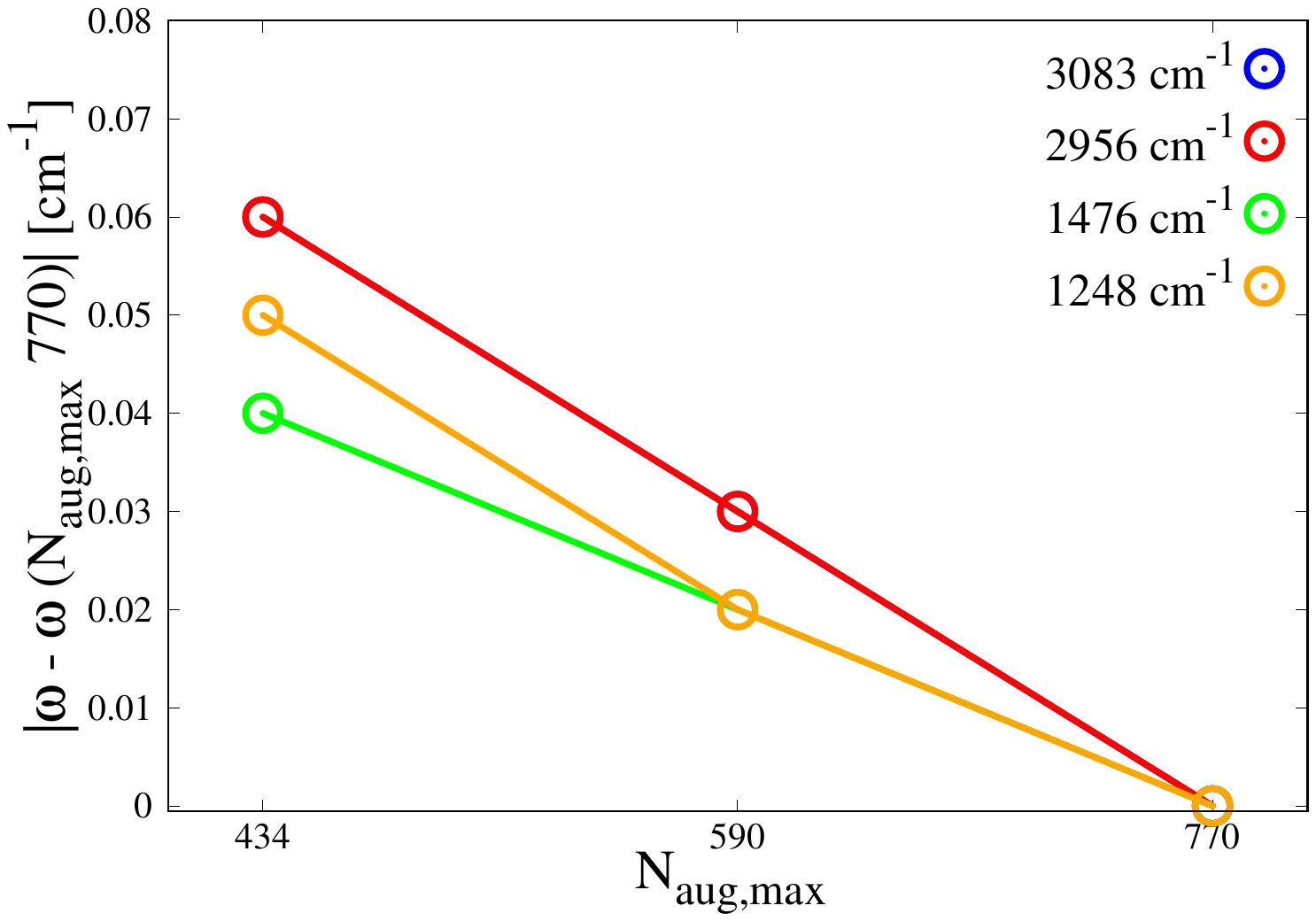}
 \caption{
 Convergence of the infrared-active vibrational frequencies of methane with respect to the angular integration grid,
 as controlled by the parameter~$N_{ang,max}$~(see text).
 We use a ``tier 2'' basis set and $N_{r,mult}$=4 here. The benchmark values are calculated using $N_{ang,max}$=770.}
 \label{fig:ch4_aug}
\end{figure}

Fig.~\ref{fig:ch4_aug} show our convergence tests with respect to the angular integration points~($N_{ang,max}$). Similarly, we find that the computed vibrational frequencies depend only weakly on the chosen angular integration grids. The maximum absolute error of the vibrational frequencies is always smaller than $0.1$~cm$^{-1}$.

\subsection{Phonon band structure compared with experiment}
\label{subsec:phonon}
\begin{figure}
 \includegraphics[width=\columnwidth]{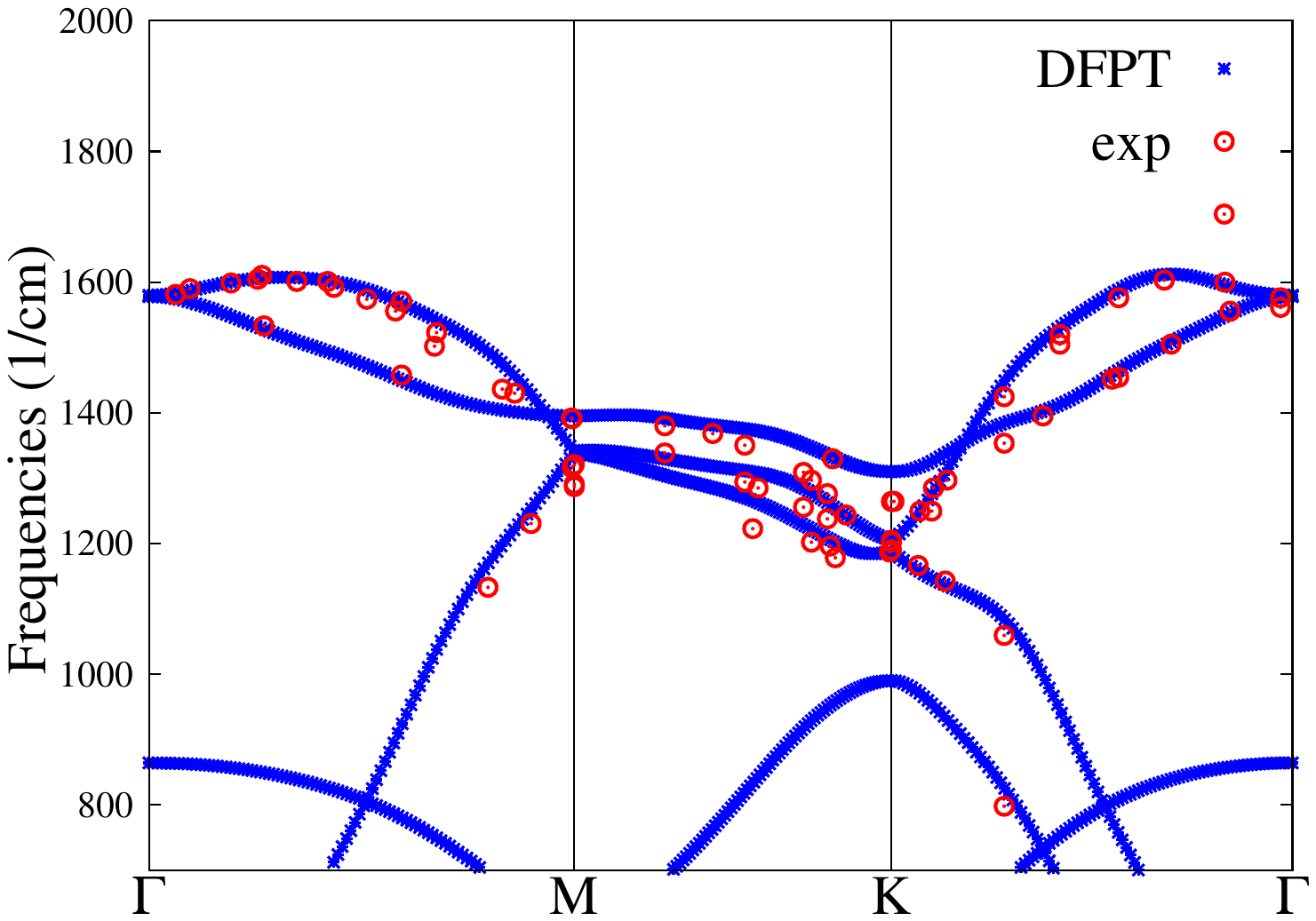}
 \caption{The graphen phonon band structure, computed at the LDA level using the DFPT method is compared with the inelastic x-ray scattering experiment data~(red points)\cite{Maultzsch2004}. }
 \label{fig:graphen}
\end{figure}

We have also validated our implementation against the experiment result. Here we use graphene as an example. The DFPT calculation has been performed with 11$\times$11$\times$1~$\mathbf{k}$-points in the primitive unit cell, real-tight settings, the ``tier 1'' basis set, and the LDA functional. 
Here the LDA is adopted because as noted in Ref.\cite{He2014}, when making the assessment of the validity of various XC functionals for computing the phonon band structure, the LDA is the functional that performs the best when compared with the experiment. In this work, we indeed find an excellent agreement between our DPFT implementation and the experiment results\cite{Maultzsch2004}, as shown in Fig.~\ref{fig:graphen}. We have also compared our DFPT results with the finite-difference calculations, which also gives an excellent agreement\cite{Shang2017}.

\section{Conclusions} 
\label{sec:conclusions}
In this work, we have shown the moving-grid effect~(atom-centered grid points moving with the atom and the weight derivatives) in second-order derivatives (i.e. vibrational frequencies) calculations with numeric atom-centered orbitals, the formulas of
moving-grid effect in Hellmann-Feynman Hessian and Pulay Hessian have been derived and implemented. In particular, we have
shown the moving-grid influence with respect to the numerical parameters used in the computation, i.e. grid, basis set, elements. 
Also, we have demonstrated that the computed vibrational frequencies by considering moving-grid effect are essentially equal to the ones obtained from finite differences. Furthermore, we have shown by considering 
acoustic sum rule, we can get the right vibrational frequencies by using the off-diagonal Hessians. 

\section{Acknowledgments}
This work was supported by the Strategic Priority Research Program of Chinese Academy of Sciences~(Grant No. XDC01040100). The author acknowledges Professor Patrick Rinke for many inspiring discussions and is grateful to Professor Matthias Scheffler for his generous support on this project.

%%%%%%%%%%%%%%%%%%%%%%%%%%%%%%%%%%%%%%%%%%%%%%%%%%%%%%%%%%%%%%%%%%
%                        Bibliography                            %
%%%%%%%%%%%%%%%%%%%%%%%%%%%%%%%%%%%%%%%%%%%%%%%%%%%%%%%%%%%%%%%%%%
\bibliography{moving_grid_effect} % Produces the bibliography via BibTeX.

\end{document}